\pdfoutput=1 
\documentclass[
superscriptaddress,
 aps,
reprint
]{revtex4-1}

\usepackage{graphicx}
\usepackage{dcolumn}
\usepackage{bm}
\usepackage{amsmath}
\usepackage{color}
\usepackage{float}
\usepackage{array,url,kantlipsum}
\usepackage{verbatim}
\usepackage{xcolor}

\newcolumntype{P}[1]{>{\centering\arraybackslash}p{#1}}

\begin{document}
\preprint{APS/123-QED}

\title{Microwave-free vector magnetometry with nitrogen-vacancy centers along a single axis in diamond}
\author{Huijie Zheng}
\email{zheng@uni-mainz.de}
\affiliation{Johannes Gutenberg-Universit{\"a}t Mainz, 55128 Mainz, Germany}
\author{Zhiyin Sun}
\email{zhiyisun@uni-mainz.de}
\affiliation{Laboratory for Space Environment and Physical Sciences, Harbin Institute of Technology, 150001 Harbin, China}
\affiliation{Helmholtz Institut Mainz, 55099 Mainz, Germany}
\author{Georgios Chatzidrosos}
\affiliation{Johannes Gutenberg-Universit{\"a}t Mainz, 55128 Mainz, Germany}
\author{Chen Zhang}
\affiliation{Institute of Physics, University of Stuttgart and Institute for Quantum Science and Technology IQST, 70174 Stuttgart, Germany}
\author{Kazuo Nakamura} 
\affiliation{Application Technology Research Institute, Tokyo Gas Company, Ltd., Yokohama, 230-0045 Japan}
\author{Hitoshi Sumiya} 
\affiliation{Advanced Materials Laboratory, Sumitomo Electric Industries, Ltd., Itami, 664-0016 Japan}
\author{Takeshi Ohshima} 
\affiliation{Takasaki Advanced Radiation Research Institute, National Institutes for Quantum and Radiological Science and Technology, Takasaki, 370-1292, Japan }
\author{Junichi Isoya}
\affiliation{Faculty of Pure and Applied Sciences, University of Tsukuba, Tsukuba, 305-8573 Japan}
\author{J\"{o}rg Wrachtrup}
\affiliation{Institute of Physics, University of Stuttgart and Institute for Quantum Science and Technology IQST, 70174 Stuttgart, Germany}
\author{Arne Wickenbrock}
\affiliation{Johannes Gutenberg-Universit{\"a}t Mainz, 55128 Mainz, Germany}
\affiliation{Helmholtz Institut Mainz, 55099 Mainz, Germany}
\author{Dmitry Budker}
\affiliation{Johannes Gutenberg-Universit{\"a}t Mainz, 55128 Mainz, Germany}
\affiliation{Helmholtz Institut Mainz, 55099 Mainz, Germany}
\affiliation{Department of Physics, University of California, Berkeley, CA 94720-7300, USA}
\affiliation{Nuclear Science Division, Lawrence Berkeley National Laboratory, Berkeley, CA 94720, USA}
 \date{\today}

\begin{abstract}
Sensing vector magnetic fields is critical to many applications in fundamental physics, bioimaging, and material science. Magnetic-field sensors exploiting nitrogen-vacancy (NV) centers are particularly compelling as they offer high sensitivity and spatial resolution even at nanoscale. Achieving vector magnetometry has, however, often required applying microwaves sequentially or simultaneously, limiting the sensors' applications under cryogenic temperature. Here we propose and demonstrate a microwave-free vector magnetometer that simultaneously measures all Cartesian components of a magnetic field using NV ensembles in diamond. 
In particular, the present magnetometer leverages the level anticrossing in the triplet ground state at 102.4\,mT, allowing the measurement of both longitudinal and transverse fields with a wide bandwidth from zero to megahertz range. 
Full vector sensing capability is proffered by modulating fields along the preferential NV axis and in the transverse plane and subsequent demodulation of the signal. This sensor exhibits a root mean square noise floor of $\approx$ 300\,pT\,/$\sqrt[]{Hz}$ in all directions. The present technique is broadly applicable to both ensemble sensors and potentially also single-NV sensors, extending the vector capability to nanoscale measurement under ambient temperatures.
\end{abstract}

\pacs{Valid PACS appear here}
\maketitle

\section{Introduction}

Sensitive vector magnetometers are exploited in applications including magnetic navigation\,\cite{cochrane2016vectorized}, magnetic anomaly detection\,\cite{lenz2006magnetic}, current and position sensing\,\cite{lenz2006magnetic}, and measuring biological magnetic fields\,\cite{RevModPhys.65.413,le2013optical}. Several versatile magnetometry platforms have emerged over the past decades, such as Hall probes, flux-gate, tunneling-magnetoresistance\,\cite{LUONG2017297}, superconducting quantum interference device (SQUID) based magnetometry\,\cite{vectormagnetometer2} and vapor cell based magnetometry\,\cite{unshieldedvectormagnetometer,PhysRevLett.113.013001}. Particularly compelling are sensors based on negatively charged nitrogen-vacancy (NV) centers in single-crystal diamond, providing high-sensitivity magnetic sensing and high-resolution imaging\,\cite{Cavity2,PhysRevB.87.155202,PhysRevApplied.8.044019}. There is growing interest in magnetic-field sensors with high spatial resolution, for example to study biological processes or the composition of materials. Utilizing NV centers for magnetometry allows measuring magnetic fields at ambient temperature and microscopic scales, providing new tools for probing various phenomena including magnetism in condensed matter systems\,\cite{pelliccione2016scanned}, semiconductor materials\,\cite{gong2017discovery} and metallic compounds\,\cite{bonilla2018strong}, and elucidating spin order in magnetic materials\,\cite{gross2017real}.


To date, diamond-based vector magnetometers have been realized by interrogating ensembles of NV centers along multiple crystallographic axes\,\cite{PhysRevApplied.10.034044,zhang2018vector} or relying on a hybrid magnetometry platform consisting of an electronic NV sensor and a nuclear spin qubit at particular positions\,\cite{PhysRevLett.122.100501}. These techniques, however, are all based on using the optically detected magnetic resonance (ODMR) technique with the requirement of applying microwaves sequentially or simultaneously\,\cite{wang2015high,zhang2018vector,PhysRevApplied.10.034044}. The requirement of microwave control brings the possibility of spurious harmonics within the measurement and hinders applications in areas where it is inherently difficult to achieve such control or where the application of microwaves is prohibitively invasive. Although NV-based sensors have been successfully implemented as vector magnetic probes at room temperature, it has remained an outstanding challenge to extend the vector capability to cryogenic temperatures (less than 4\,K) due to difficulties of thermal management. The heat from the applied microwaves is unavoidable and causes temperature variations, restricting the sensors for numerous innovative applications, such as mapping the magnetization of individual atomic layers of van der Waals materials\,\cite{thiel2019probing}.

We propose and demonstrate a protocol that enables vectorial measurement of magnetic fields by interrogating an ensemble of NV centers aligned along only a single crystallographic axis (we do not use a preferential-orientation NV-diamond sample) at the ground-state level anticrossing (GSLAC). By applying two orthogonal alternating fields, along and perpendicular to the chosen axis, our technique offers direct and simultaneous readout of all three magnetic components, free from systematic errors during reconstruction. In contrast to existing methods, our approach does not employ microwave fields. Thus it is possible to extend NV-based vector magnetic sensing techniques to cryogenic temperatures, representing an important advance in magnetometry.

The method can be potentially extended to single-NV probes. This will facilitate extraction of complete vector information of the magnetic field to be measured with nanoscale spatial resolution under ambient temperatures. A nanoscale \textit{vector} magnetometer would bring a wealth of additional information and is motivated by numerous applications, e.g., noninvasive tracking of particle motion in intracellular medium\,\cite{natnanotech63582011,Chen_2013} and discerning the directionality of action-potential firing\,\cite{PhysRevLett.122.100501}. The single-NV vector magnetic-field probe, in some cases, can address the problem of microscopic characterization of novel spin textures\,\cite{dovzhenko2018magnetostatic}, which, in the absence of vector information, would rely on system-dependent assumptions, artificially restricting the manifold of solutions compatible with experimental results.

\begin{figure*}
  \centering
\includegraphics[width=\textwidth]{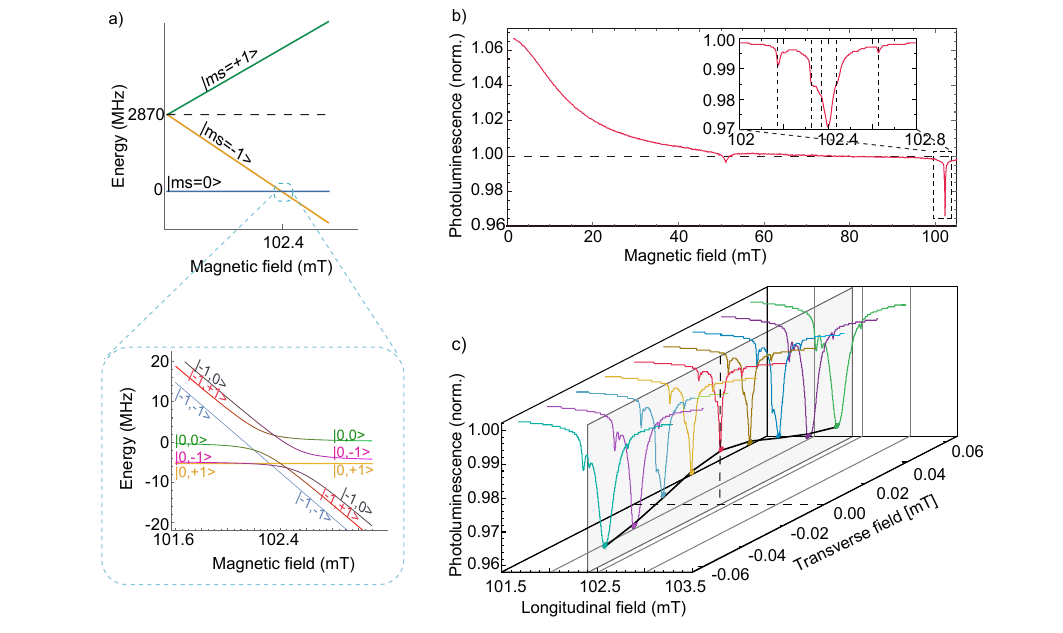}
\caption 
{ \label{Figure1} 
(a) The ground-state energy level scheme of the NV center as a function of the applied axial field. The energy levels either cross or do not cross depending on the mixing between them, depicted in detail in the inset. (b) The PL signal as a function of the applied axial magnetic field, normalized to their respective signals at 80\,mT. The inset shows a detailed view around the GSLAC trace. Features at around 51\,mT, due to cross relaxations\,\cite{Wood2016_2,hall2016detection} and possibly excited-state LAC\,\cite{Ivanov2016}, have been extensively investigated. (c) Traces of the PL signal around the GSLAC under various transverse fields. The amplitude of the contrast extracted from the curves is shown as a two-dimensional plot in a plane, indicated by solid dots in corresponding trace colors.}
\label{fig:fullscan}
\end{figure*}
\section{Magnetometry Method}
This microwave-free technique for magnetic sensing is based on detecting changes in NV-photoluminescence (PL) under optical pumping near the GSLAC. It was proposed and demonstrated for sensing the longitudinal component of a magnetic field in \citet{wickenbrock2016microwave} and non-quantitatively indicating a transverse component by T1 relaxometry detection\,\cite{PhysRevApplied.6.064001}. As both the longitudinal and transverse magnetic fields can lead to a change in the PL signal at the GSLAC and the response to the direction of the transverse component is highly non-trivial\,\cite{NatureSingleNV1,levelanticrossing}, achieving vectorial sensing of magnetic fields with a single NV center or an ensembles along a single crystallographic axis faces a number of significant challenges.

In the following, we address the challenges of measuring all Cartesian components simultaneously and precisely.
The NV center is an atomic-scale defect consisting of a substitutional nitrogen adjacent to a vacancy in the diamond lattice. It has a spin-triplet ground state ($\mathbf{S}$=1), which can be optically polarized to $|m_s=0\rangle$ and read out due to a spin-dependent intersystem crossing into an intermediate singlet state.
Without magnetic field the $|m_s=\pm 1\rangle$ states are (nearly) degenerate; however, owing to spin-spin interaction, these states lie higher in energy than the $|m_s= 0\rangle$ state. This is the so-called zero-field splitting $\mathcal{D}$ between the magnetic sublevels corresponding to a frequency difference of $2.87$\,GHz. Brought to degeneracy via the Zeeman effect, a subset of NV centers' magnetic sublevels experience a complex GSLAC at an axial field $B_{z} \approx 102.4$\,mT\,\cite{wickenbrock2016microwave,PhysRevApplied.6.064001}. Figure\,\ref{fig:fullscan}\,(a) shows the energy levels of the NV center as a function of an applied magnetic field including the coupling to the intrinsic nuclear spin of nitrogen ($\mathbf{I}$=1). 

Figure\,\ref{fig:fullscan}\,(b) shows the PL signal as a function of the axial magnetic field with zero transverse fields. A remarkably sharp feature around 102.4\,mT, zoomed-in in the inset, indicates the GSLAC. In the inset, several additional features are visible which can be attributed to cross-relaxation with the nearby spin bath\,\cite{wickenbrock2016microwave,hall2016detection,armstrong2010nv,AcostaPRB2010}. A detailed study of these features is currently being conducted and will be presented in a separate manuscript. 
Transverse fields couple the $|m_s=0\rangle$ and $|m_s=-1\rangle$ magnetic sublevel manifold and therefore affect contrast and amplitude of the GSLAC feature. Traces of the GSLAC feature for several transverse fields in the range of $\pm$0.06\,mT, are depicted in Fig.\,\ref{fig:fullscan}\,(c). The amplitudes of the GSLAC feature as a function of transverse field is indicated by the trace-colored dots and connected with the black line.  In summary, the GSLAC contrast exhibits a relatively narrow (FWHM$\approx$\,38\,$\mu$T) magnetic resonance feature as a function of transverse magnetic field centered around zero transverse field. 

In order to describe the vector sensing mechanism, we first analyze the Hamiltonian of the triplet ground state around the GSLAC.
The system can be modelled by only considering $|m_s=0\rangle$ and $|m_s=-1\rangle$. The $|m_s=+1\rangle$ is ignored in the following because it is at a much higher energy than the $|m_s=0\rangle$ state which is preferentially populated under optical excitation\,\cite{NatureSingleNV1}. With the hyperfine interaction between the NV electron spin and the nuclear spin of the intrinsic nitrogen atom (for details see \cite{auzinsh2018hyperfine}), the Hamiltonian is expressed in the basis of \{$m_s,m_I$\}. 
Here we write a two-level Hamiltonian in the subspace \{$|0,+1\rangle,|-1,+1\rangle$\} since the spins are efficiently polarized to the $|0,1\rangle$ state for a $^{14}$N-V center\,\cite{PhysRevApplied.6.064001} under optical excitation. For a given NV center, we define the $z$-axis along the symmetry axis of the center. In the presence of an arbitrary magnetic field $\mathbf{B}=(\mathrm{B}_x,\mathrm{B}_y,\mathrm{B}_z)$ and neglecting the nuclear Zeeman shift, the reduced Hamiltonian is given by 
\begin{equation}
\mathcal{H}_r=
\begin{pmatrix}
    0 &\gamma_e\mathrm{B}_\bot\frac{e^{-i\phi}}{\sqrt{2}} \\
    \gamma_e\mathrm{B}_\bot\frac{e^{+i\phi}}{\sqrt{2}} &  \mathcal{D}-\gamma_e\mathrm{B}_z \\
\end{pmatrix},\label{eq:eqA1}
\end{equation}
where $\mathrm{B}_\bot$ is the transverse magnetic field ($|\mathrm{B}_\bot|=\sqrt{\mathrm{B}_x^2+\mathrm{B}_y^2}$), $\phi$ is the angle defined by tan $\phi=\mathrm{B}_y/\mathrm{B}_x$, and $\gamma_e$ is the electron gyromagnetic ratio. 
In the absence of transverse fields, the $|0,+1\rangle$ state does not mix with any other states [see Fig.\,\ref{fig:fullscan}\,(a)]. Therefore, if the center is fully polarized to this state, the PL should not depend on the exact value of the longitudinal magnetic field. However, in the presence of the spin bath producing randomly fluctuating magnetic fields, there arises an effective coupling between the eigenstates resulting in depolarization of the NV center and a corresponding drop in PL near the level crossings. 

We introduce the axial field difference from the crossing, $\gamma_e \delta \mathrm{B}_z=\mathcal{D}-\gamma_e\mathrm{B}_z$. If $|\delta \mathrm{B}_z|\gg|\mathrm{B}_\bot|$, far from the avoided crossing region, the PL is insensitive to the transverse field. Conversely, if $|\mathrm{B}_\bot|\gg|\delta \mathrm{B}_z|$, the signal becomes insensitive to small changes in the longitudinal field. In other words, near the GSLAC, the PL can be used to determine the transverse and longitudinal components of the magnetic field to be measured. This magnetic-vector sensing protocol can be extended to single-NV probes, and therefore, nanoscale sensing volume, since it just relies on intrinsic properties of the NV center and the presence of a spin bath.

\begin{figure}
\centering
\includegraphics[width=\columnwidth]{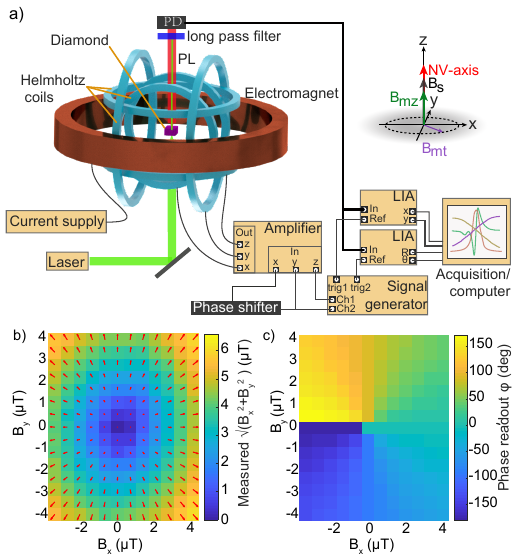}
\caption{(a) Experimental setup for the microwave-free magnetometer and the set Cartesian coordinate system. The diamond is placed in the center of a 3-D Helmholtz coil pairs. A static magnetic field ($B_s$) is applied along a NV axis (noted as $z$ direction) by a customized electromagnet. The static magnetic field, modulating fields ($B_{mz}$ along $z$ and $B_{mt}$ in the transverse plane)and the axis of NV centers in the NV frame coordinate system are displayed. 
The measured magnitude (b) and the direction angle (c) as a function of the Cartesian components of the applied field. The reconstructed field vectors are also shown in (b), depicted by the red arrows.
}
\label{fig:nvsetup}
\end{figure}

Based on the Hamiltonian (Eq.\,\ref{eq:eqA1}) and the assumption of an isotropic spin bath, we expect that the effect of a transverse magnetic field on the intensity of PL should not depend on the direction of the transverse field. In fact, we observe this experimentally, see Fig.\,\ref{fig:nvsetup} (b). The PL, however, does depend on the magnitude of the applied transverse field. Therefore, we have a sensor for the magnitude of the transverse field. Similar to how it is possible to measure the field vector with a scalar magnetometer by applying modulated fields in different directions, it is also possible, as we demonstrate here, to measure both Cartesian components of the transverse field with our sensor.

\begin{figure*}
\centering
\includegraphics[width=\textwidth]{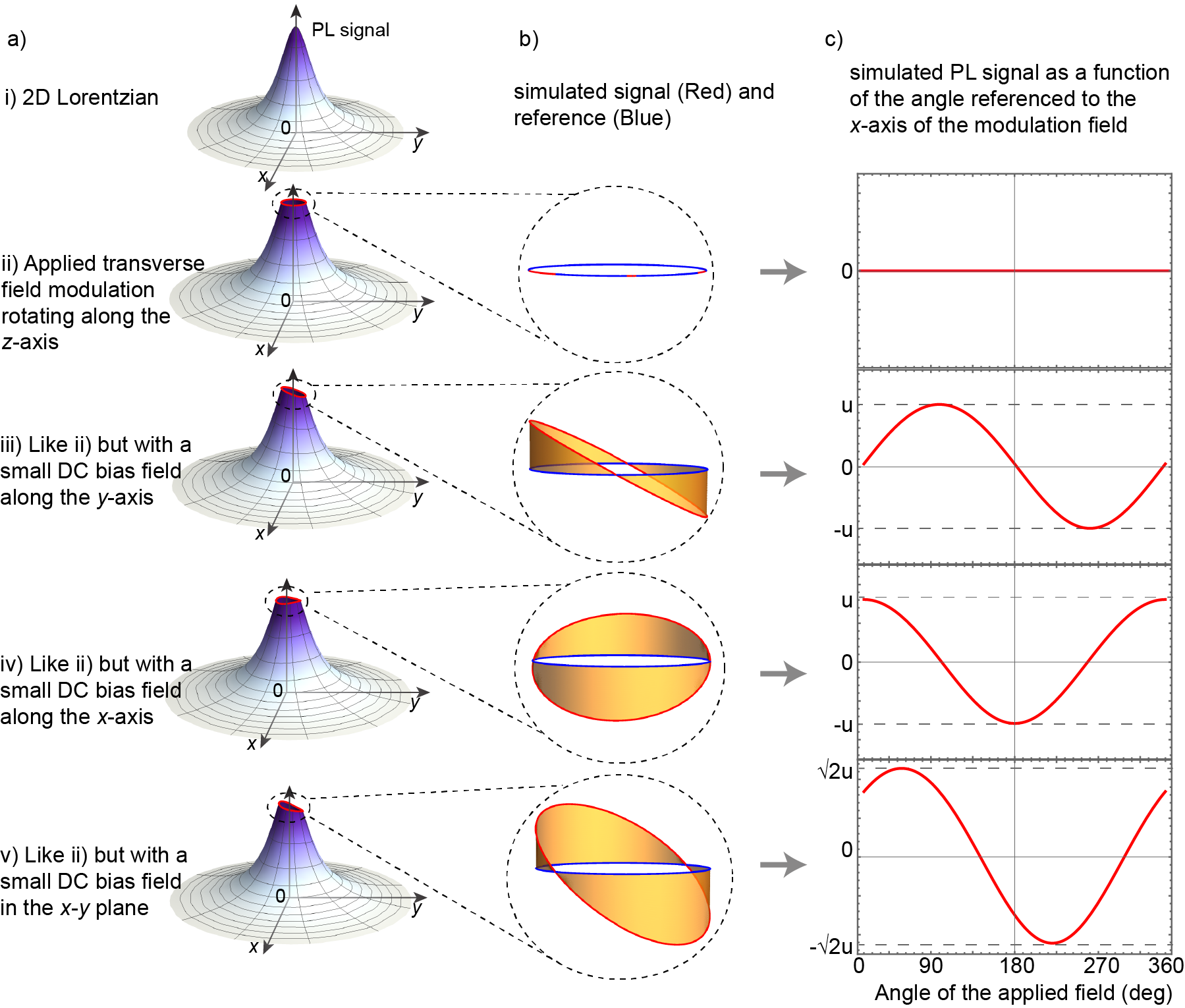}
\caption{(a) Simulated PL signals as a function of transverse field. The red curves are the trajectories of PL signals superimposed on the 2-D Lorentzian contrast function. Figures i) to v), correspond to different show cases: i) no modulated fields and no bias fields, ii) with modulated fields but no bias fields (the time-averaged PL drops), iii) with modulated fields and bias field along $y$, iv) with modulated fields and bias field along $x$, v) with modulated fields and bias field in $x$-$y$ plane, respectively. (b) The simulated PL signals with modulating magnetic field in the presence/absence (red curve/blue curve) of a bias field. (c) Simulated PL signals [in the same coordinate system as in column (a)] as a function of the angle of the modulated field referenced to the $x$-axis for transverse fields, flipped by 180$^{\rm{o}}$ corresponding to (a) and (b).}
\label{fig:xymethod}
\end{figure*}

A typical method to adapt a scalar magnetometer for vector measurement is to apply mutually orthogonal fields modulated at different frequencies. Thus it is possible to determine the components along each direction by individually demodulating the signal\,\cite{PhysRevLett.113.013001}. In this work we propose a method to realize vector-field sensing in the $x$-$y$ plane using a transverse field rotating around the $z$-axis [Fig.\,\ref{fig:nvsetup}\,(a)] with just one frequency. 
To gain an intuitive understanding, we approximate the PL lineshape as a function of transverse magnetic fields with a 2-D Lorentzian centered around $\mathrm{B}_x=\mathrm{B}_y=0$, Fig.\,\ref{fig:xymethod} (a) (i). With a transverse field applied that is rotating around $z$, the PL signal will be reduced but remains unmodulated, indicated by the red curve in Fig.\,\ref{fig:xymethod} (a) (ii). In the presence of an additional static transverse field, the PL signal shows a modulation at the rotation frequency with a minimum when the rotating field points in the same direction as the field under interrogation and a maximum when both are antiparallel, shown in Fig.\,\ref{fig:xymethod} (a) (iii), (iv) and (v). The difference between the PL signals with (red curve) and without (blue curve) applied transverse field is shown in Fig.\,\ref{fig:xymethod} (b). This is then demodulated by a Lock-in amplifier (LIA) which delivers the information of both the amplitude and the angle of the magnetic field to be measured, as shown in Fig.\,\ref{fig:xymethod} (c). The reference phase of the LIA was set so that a magnetic field along the $x$-axis corresponds to phase zero (and negative amplitude). The LIA output shows a maximum value at $0^{\rm{o}}$ when applying a field along $x$-axis, shown in Fig.\,\ref{fig:xymethod} (c). An applied field in any other direction leads to an oscillating PL signal with a corresponding phase. Therefore, the phase output of the LIA is the angle between the transverse field to be measured and the defined $x$ and $y$ axes.

In addition to the longitudinal magnetic field measurement\,\cite{wickenbrock2016microwave}, all Cartesian magnetic-field components can be directly read out in real time with equal sensitivity in all directions. Note here, that the reference phase of the LIA for the transverse-field signal demodulation sets the coordinate axes in the $x$-$y$ plane while the phase for $z$-axis demodulation is tuned to maximize the amplitude of the response signal. 

\section{Vector-sensing Demonstration}

The experimental apparatus includes a custom-built electromagnet and three pairs of orthogonal Helmholtz coils wound on a 3-D printed mount. The electromagnet can be moved with a computer-controlled 3-D translation stage (Thorlabs PT3-Z8) and a rotation stage (Thorlabs NR360S, $x$-axis). The NV-diamond sensor is placed in the center of both the magnetic bore and three pairs of integrated orthogonal Helmholtz coils. The diamond can be rotated around the $z$-axis. 
This provides all degrees of the necessary freedom for placing the diamond in the center of the magnet and aligning the NV axis parallel to the magnetic field. 

A two-channel function generator (Tektronix AFG 3022A) provides sinusoidal signals for field modulations in the longitudinal and transverse directions and references for the demodulation by two LIAs. The signal from one of the channels is split in two with one of them passing through a phase shifter. These two signals with the same frequency but 90-degree-shifted relative phase are applied to two pairs of the Helmholtz coils (along the $x$ and $y$ axes). All three signals are amplified via a homemade 3-channel current amplifier before reaching the Helmholtz coils.

The light source is a solid-state laser emitting at a wavelength $\lambda$\,=\,532\,nm (laser Quantum Gem 532). The PL emitted by the diamond sample is collected with a parabolic lens and detected with a photodetector (Thorlabs PDA 36A).

The sensor is a 99.97\% $^{12}$C, (111)-cut diamond single crystal, with dimensions $0.71\,\mathrm{mm}\times 0.69\,\mathrm{mm}$ and a thickness $0.43\,\mathrm{mm}$. It was laser-cut from a $^{12}$C-enriched diamond single crystal grown by the temperature gradient method at high pressure (6.1\,GPa) and high temperature ($1430^{\mathrm{o}}$C). A metal solvent containing a nitrogen-getter and carbon powder prepared by pylolysis of 99.97\% $^{12}$C-enriched methane as a carbon source were used\,\cite{Nakamura2007}. It was irradiated with 2\,MeV electrons from a Cockcroft-Walton accelerator to a total fluence of $1.8\times10^{18} \mathrm{cm}^{-2}$ at room temperature, and annealed at 800$^{\mathrm{o}}$C for 5 hours. This source diamond was reported to have 3-ppm initial nitrogen and 0.9-ppm NV$^-$ after conversion, measured by electron paramagnetic resonance techniques\,\cite{Wolf2015}. This diamond provides remarkable narrow GSLAC-features with small residual couplings to $^{13}$C nuclear spins which is essential for the sensitivity of the proposed method.

\begin{figure}
\centering
\includegraphics[width=\columnwidth]{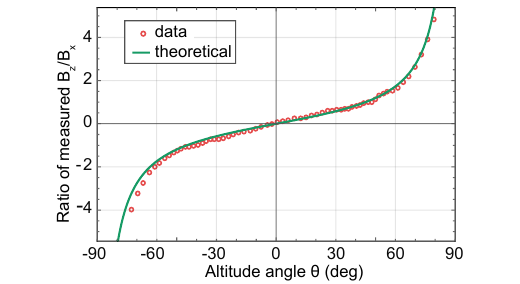}
\caption{The ratios of measured fields $B_z/B_x$ corresponding to different angles are calculated from the applied fields in $x$-$z$ plane. The solid green line shows an arctangent dependence.}
\label{fig:xzdemonstration}
\end{figure}

The Helmholtz coil pairs for the field modulation can also be used to calibrate the response to AC and DC magnetic fields  $\mathbf{B}=(\mathrm{B}_x,\mathrm{B}_y,\mathrm{B}_z)$. The applied fields, in the range of $\pm$4\,$\mu$T along each direction, are calibrated by flux gate magnetometers and consistent with a \textit{priori} calculations from the known coil geometry and the applied currents. The single-frequency-modulation 2-D vector magnetometry method is demonstrated by mapping the amplitudes and phase of the LIA output as a function of the applied fields in the $x$-$y$ plane, see Fig.\,\ref{fig:nvsetup} (b) and (c).
We also infer the field direction in $x$-$z$ plane from measurements of the ratio $\mathrm{B}_z$/$\mathrm{B}_x$, see Fig.\,\ref{fig:xzdemonstration}.

The measurement in the $z$-axis is carried out by applying an additional modulated field along $z$, modulated with a different frequency (than the one in the transverse direction) and the corresponding demodulation of the PL signal. A matrix of 2-D vector magnetic fields in the $x$-$z$ plane was measured to demonstrate this method. The angle of the field can be calculated as $\arctan({\mathrm{B}_z}/{\mathrm{B}_x})$ from the applied currents and calibration factors of the two coils. Figure\,\ref{fig:xzdemonstration} shows that the ratios of the measured fields $\mathrm{B}_z$ and $\mathrm{B}_x$ correspond to different angles following the expected arctangent curve. 

\begin{figure*}
\centering
\includegraphics[width=\textwidth]{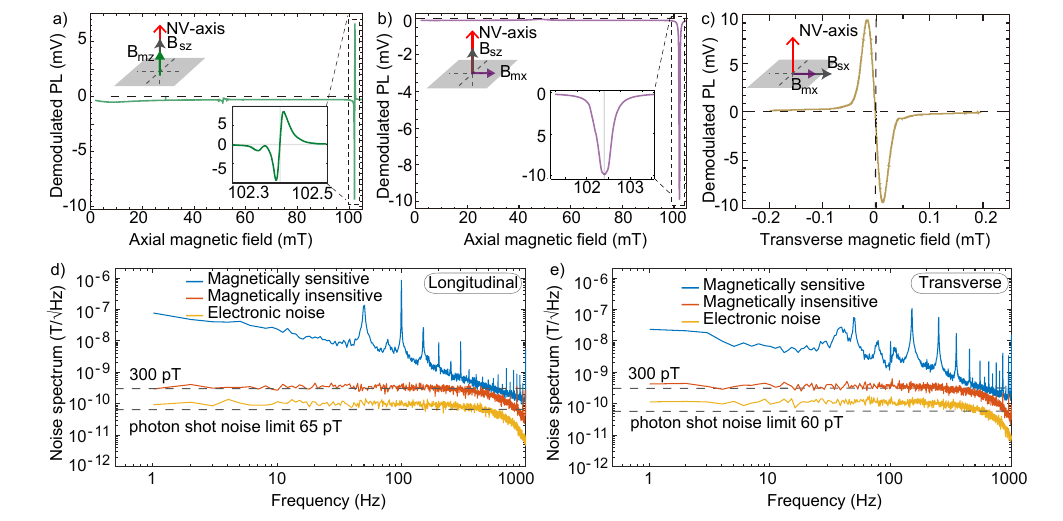}
\caption{(a) Demodulated PL signal, LIA output X, the measure of magnetic field sensitivity, as a function of axial magnetic field ($B_{sz}$) with a small added modulation ($B_{mz}$). (b) Demodulated PL signal as a function of axial magnetic field with an added transverse magnetic field modulation ($B_{mx}$). (c) Demodulated PL signal as a function of transverse field ($B_{sz}$) along the $x$-axis while modulating the magnetic field along the same direction as (b). (d) Longitudinal ($z$-axis) magnetic field noise spectrum. The blue line indicates the noise in the magnetically sensitive configuration at a magnetic field of 102.4\,mT, the red line indicates a noise in the magnetically insensitive configuration (average noise between 1$-$500\,Hz is 300\,pT/$\sqrt{\rm{Hz}}$), and the amber line illustrates the electronic noise (average noise between 1$-$500\,Hz is 100\,pT/$\sqrt{\rm{Hz}}$). The decrease in signal for frequencies above 1\,kHz is due to the filtering of the LIA. Photon shot noise is estimated as 65\,pT/$\sqrt{\rm{Hz}}$. (e) Transverse magnetic field noise spectrum ($x$-axis). The blue line indicates the noise in the magnetically sensitive configuration measured at zero transverse field, the red line indicates the noise in the magnetically insensitive configuration at a transverse magnetic field of 0.2\,mT (average noise between 1$-$500\,Hz is 300\,pT/$\sqrt{\rm{Hz}}$), and amber line shows the electronic noise spectrum (average noise between 1$-$500\,Hz is 100\,pT/$\sqrt{\rm{Hz}}$). Photon shot noise is estimated as 60\,pT/$\sqrt{\rm{Hz}}$. }
\label{fig:noisefloor}
\end{figure*}

Before full vector-sensing protocol is demonstrated, the sensitivity along both the longitudinal and transverse directions ($z$ and $x$) as well as possible cross-talk effects were evaluated. Derivatives of the fluorescence signals in Fig.\,\ref{fig:fullscan}\,(b), detected in the properly phased LIA X output while applying sinusoidally modulating fields along $z$-axis or $x$-axis in the presence of a static field along $z$-axis, are shown in Fig.\,\ref{fig:noisefloor}\,(a) and (b). The modulation frequencies were 3.7\,kHz and 2.3\,kHz, respectively, and the modulation depth $\approx$20\,$\mu$T. Pronounced magnetically dependent features around the GSLAC were detected. In the case of $z$-axis modulation, the resulting demodulated PL signal depends linearly on the magnetic field in the region near the GSLAC [Fig.\,\ref{fig:noisefloor}\,(a)], while it is first-order insensitive to the $z$-axis magnetic field when applying modulation along the $x$-axis, [Fig.\,\ref{fig:noisefloor}\,(b)]. This demonstrates the absence of crosstalk between the different modulating fields in the proximity of the GSLAC. At a longitudinal field corresponding to the GSLAC and with an applied $x$-axis modulation the demodulated PL signal shows a linear dependence on the $x$-axis magnetic field [see Fig.\,\ref{fig:noisefloor} (c)]. 

Figures\,\ref{fig:noisefloor} (d) and (e) show the single channel magnetic sensitivity along the $z$ and $x$ axes, respectively. These are calibrated by linearly fitting the data near the zero-crossing of the corresponding derivative curves [Fig.\,\ref{fig:noisefloor} (a) and (c)] and the slope of this line is used to translate the LIA output signal to magnetic field. For noise measurements, the LIA output is recorded for 1\,s while the background magnetic field is set to a point where the LIA X is zero. The data are fast-Fourier transformed and show noise floor in different configurations [see Fig.\,\ref{fig:noisefloor}\,(d) and (e)], and thus the sensitivity, for a given bandwidth. Despite the dominant 1/f noise (near an order of magnitude higher in the longitudinal direction presumably due to the power supply of the electromagnet), the device exhibits a noise floor of around 300\,pT/$\sqrt{\mathrm{Hz}}$ in both the longitudinal direction and the measured transverse direction. The noise for the magnetically insensitive configuration in both cases was measured at magnetic fields of 0.2\,mT off the GSLAC field. The electronic noise floor ($\approx100$\,pT/$\sqrt{\rm{Hz}}$) was measured by turning off the green excitation light and acquiring the output of the LIA. The photon shot noise limit for the two measurements is 65\,pT/$\sqrt{\mathrm{Hz}}$ and 60\,pT/$\sqrt{\mathrm{Hz}}$, respectively, which are calculated based upon the number of incident photons at the photodetector, the contrast and width of the magnetically sensitive feature in the GSLAC spectral response.
The overall noise is dominated by environmental noise. The noise can be further suppressed deploying a differential detection schemes\,\cite{Wolf2015}.

\begin{figure*}
\centering
\includegraphics[width=\textwidth]{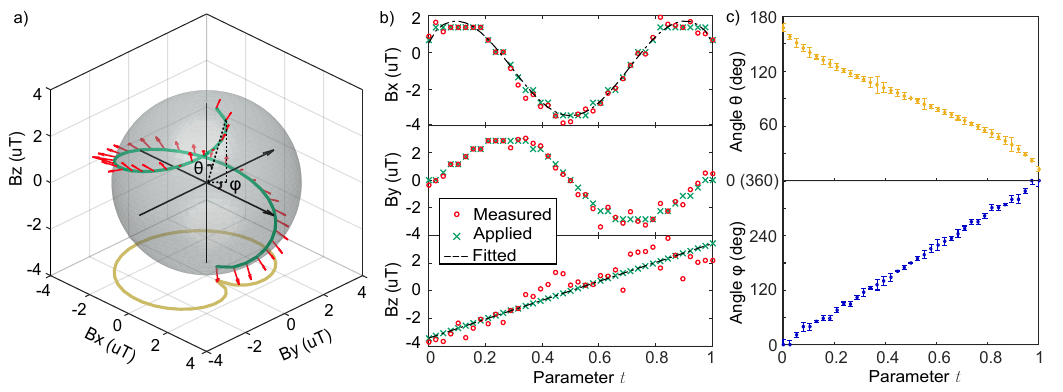}
\caption{Demonstration of full vector capability. (a) Trajectory of the detected magnetic fields using the microwave-free vector magnetometer. The green curve indicates the 3-D applied field and the brown curve is the projection on $x$-$y$ plane. The red arrows represent the vectors of the measured fields. (b) The three Cartesian components of both the applied (green points) and measured (red circles) magnetic fields for each point. The applied field follows a parametric curve (black dashed lines) with $\mathrm{B}_x=\sqrt{|\rm{B}|^2-Bz^2}\,\mathrm{cos}(2\pi t)$, $\mathrm{B}_y=\sqrt{|\rm{B}|^2-Bz^2}\,\mathrm{sin}(2\pi t)$ and $\mathrm{B}_z=6.82\,t-3.41$. (c) The altitude angle (between $\mathbf{B}$ and the $z$-axis) $\theta=\mathrm{arccos}(\rm{B}_z/|\rm{B}|)$ (yellow dots) and the azimuth angle (between the projection of $\mathbf{B}$ in the $x$-$y$ plane and the $x$-axis) $\phi=\mathrm{arctan}(\rm{B}_y/\rm{B}_x)=2\pi t$ (blue dots) for each measured point. In the experiment, the altitude angle $\theta$ decreases in time from $180^o$ to $0^o$ and the azimuth angle $\rm{\phi}$ increases from $0^o$ to $360^o$.}
\label{fig:3dvectordemonstration}
\end{figure*} 
As a demonstration of full-vector sensing capacity, a set of static magnetic field vectors, the trajectory of which was designed along a 3-D a spiral curve on a sphere, was applied and measured. The applied field value curve is shown in Fig.\,\ref{fig:3dvectordemonstration}\,(a) and the corresponding amplitudes along each coordinate axis are displayed in Fig.\,\ref{fig:3dvectordemonstration}\,(b). The component in $z$ direction is $|\rm{B}|\cos\theta$, where $|\rm{B}|$ is the magnitude of the applied magnetic field vectors and $\theta$ is the altitude angle (between the magnetic field to be measured and the $z$-axis). The $\mathrm{B}_x$ and $\mathrm{B}_y$ are $|\rm{B}|\sin\theta\cos\phi$ and $|\rm{B}|\sin\theta\sin\phi$, respectively, where $\rm{\phi}$ is the azimuth angle (between the projection of $\mathbf{B}$ in the $x$-$y$ plane and the $x$-axis). This corresponds to the values of $\mathrm{B}_x$ and $\mathrm{B}_y$ shown in Fig.\,\ref{fig:3dvectordemonstration}\,(b). The measured field components in $x$ and $y$ directions show good agreement with the amplitudes determined by a \textit{priori} calculations. The scatter in the data can be attributed to environmental noise in the laboratory and the applied field. The trajectory was measured multiple times and the angles were reconstructed every time. Figure\,\ref{fig:3dvectordemonstration} (c) shows the average angle with the statistical error. Note here, that all the experiments were performed in a lab environment without magnetic shielding.

With the basic protocol established, this simultaneous vector magnetometry method should be extendable to single-NV probes. Single NV centers in diamond have been exploited to detect fluctuating magnetic fields (used as scalar relaxometry magnetometers) without microwaves\,\cite{PhysRevApplied.6.064001,wood2017microwave} as they share the spin dynamics near the GSLAC investigated above and in literature\,\cite{PhysRevApplied.6.064001}. Both techniques, the presently realized GSLAC-based vector magnetometry and relaxometry magnetometry, rely on monitoring the PL signal when the NVs are precisely turned to/near the GSLAC. Since they are operated with similar apparatus, the experimental setup for the latter can be extended for vector magnetic sensing by adding a set of 3-D Helmholtz coil pairs and two LIAs. As there are no technical barriers of implementing the protocol onto a single NV center, it appears realistic to achieve nanoscale vector magnetometry applicable in a broad range of temperatures including below 4\,K.  

\section{Conclusion}
In summary, we have proposed and demonstrated a sensing method allowing simultaneous recording of all three Cartesian components of a vector magnetic field using a solid-state spin sensor. The method operates at the GSLAC of NV centers, and does not employ microwaves in the measurement. Further optimization of the apparatus will allow a compact vector magnetometer well suited for geophysical field measurement or biophysical imaging. The present method can be applied to the anticrossings in other color-center systems.

The GSLAC-based vector magnetometer using NV centers along a single axis exhibits a root mean square noise floor of $\approx$\,300\,pT in a transverse and the longitudinal direction. While the technique was demonstrated without monitoring the intensity of the pump laser power, future experiments will utilize differential detection schemes and suppress laser-related noise.
In addition, combination with infrared-absorption-based readout\,\cite{Cavity2}, or enhancement by optical cavities\,\cite{PhysRevApplied.8.044019} will allow for magnetic-field sensing with a sensitivity reaching or even exceeding the PL shot-noise limit.

This technique should be extendable to single NV sensors, in this case, it can be expected to advance nanoscale real-time magnetic sensing and imaging applications, such as single molecule imaging. With the ability to sense vector magnetic fields, this method could potentially enable real-time imaging of magnetic dipoles with arbitrary orientations.



\section{Acknowledgements}
We thank O. Tretiak, J. W. Blanchard and P. Neumann for informative discussions and helpful advice. This work was supported by the EU FET-OPEN Flagship Project ASTERIQS (action 820394), and the German Federal Ministry of Education and Research (BMBF) within the Quantumtechnologien program (FKZ\,13N14439 and FKZ\,13N15064), and the Cluster of Excellence “Precision Physics, Fundamental Interactions, and Structure of Matter” (PRISMA+ EXC 2118/1) funded by the German Research Foundation (DFG) within the German Excellence Strategy (Project ID\,39083149), and the International Postdoctoral Exchange Fellowship Program with No.\,20171023, and DFG via GRK2198 and the Japan Society of the Promotion of Science (JSPS) KAKENHI (No.17H02751).





\bibliographystyle{apsrev4-2-2}
\bibliography{literature}

\end{document}